\documentclass[twocolumn]{aastex62}

\newcommand{\oxy}{\ensuremath{\mathrm{O}_2}}

\newcommand{\methane}{\ensuremath{\mathrm{CH}_4}}
\newcommand{\nitrous}{\ensuremath{\mathrm{N}_2\mathrm{O}}}

\newcommand{\snr}{\ensuremath{\mathrm{S/N}}}
\newcommand{\rp}{\ensuremath{\mathcal{R}}}
\newcommand{\dl}{\ensuremath{\Delta\lambda}}
\newcommand{\dv}{\ensuremath{\Delta v}}
\newcommand{\kms}{\ensuremath{\mathrm{km\,s^{-1}}}}
\newcommand{\ccf}[1]{\ensuremath{\mathrm{CCF_{#1}}}}
\newcommand{\Mag}{\ensuremath{\mathrm{mag}}}
\newcommand{\voff}{\ensuremath{v_\mathrm{offset}}}
\newcommand{\vsys}{\ensuremath{v_\mathrm{sys}}}
\newcommand{\rsun}{\ensuremath{R_\odot}}
\newcommand{\rstar}{\ensuremath{R_\star}}
\newcommand{\msun}{\ensuremath{M_\odot}}
\newcommand{\mstar}{\ensuremath{M_\star}}

\received{7 November 2018}
\revised{18 December 2018}
\accepted{20 December 2018}

\submitjournal{ApJL}

\shorttitle{Detectability of extraterrestrial \oxy{} with the ELTs}
\shortauthors{Serindag \& Snellen}

\begin{document}

\title{Testing the detectability of extraterrestrial \oxy{} with the ELTs using real data with real noise}

\correspondingauthor{Dilovan Serindag}
\email{serindag@strw.leidenuniv.nl}

\author[0000-0001-7987-2079]{Dilovan B. Serindag}
\affiliation{Leiden Observatory, Leiden University, Postbus 9513, 2300 RA Leiden, The Netherlands}

\author[0000-0003-1624-3667]{Ignas A. G. Snellen}
\affiliation{Leiden Observatory, Leiden University, Postbus 9513, 2300 RA Leiden, The Netherlands}

\begin{abstract}

The future extremely large telescopes (ELTs) are expected to be powerful tools to probe the atmospheres of extrasolar planets using high-dispersion spectroscopy, with the potential to detect molecular oxygen in Earth-like planets transiting nearby, late-type stars. So far, simulations have concentrated on the optical 7600\,\AA{} A-band of oxygen using synthetic noise distributions. In this paper, we build upon previous work to predict the detectability of molecular oxygen in nearby, temperate planets by using archival, time-series data of Proxima Centauri from the high-dispersion UVES spectrograph on ESO's Very Large Telescope (VLT). The brightest transiting M-dwarfs are expected to be about 25 times fainter than Proxima, a factor that is similar to the difference in light-gathering power between the VLT and the future ELTs. By injecting synthetic oxygen transmission signals into the UVES data, the \oxy{} detectability can be studied in the presence of real data with real noise properties. Correcting for the relatively low throughput ($\sim$4\%) of the Proxima spectra to an assumed 20\% throughput for a high-dispersion spectrograph on the European ELT, we find that the molecular oxygen signature of an Earth-twin transiting a nearby ($d \approx 7 \,\mathrm{pc}$) M5V star can be detected in 20--50 transits (a total of 70--175 hours of observing time). This estimate using more realistic simulations is close to previous predictions. Novel concepts that increase the instrumental throughput can further reduce the time span over which such observations need to be taken.

\end{abstract}


\section{Introduction} \label{sec:intro}

The search for biosignature gases in the atmospheres of terrestrial exoplanets will be an important component in the search for extraterrestrial life. Finding such compounds in thermodynamic disequlibrium and abundances inconsistent with abiotic processes will be suggestive of life \citep[e.g.,][]{lovelock1965,lippincott+1967,meadows2017,meadows+2018}. In particular, the simultaneous atmospheric detection of \oxy{} and a reducing gas, such as \methane{} or \nitrous{}, has been suggested as a probe for biological activity \citep[e.g.,][]{lovelock1975,sagan+1993,segura+2005} since such a combination is highly mutually-reactive and would not persist without continuous resupply. One promising method to search for molecular oxygen is to probe the spectra of temperate terrestrial exoplanets transiting late-type M-dwarfs  using the future extremely large telescopes (ELTs\footnote{We adopt ELT as a generic abbreviation for the future giant segmented mirror telescopes, and use ``European ELT" if we specifically mean the 39m telescope being built by ESO.}) at high dispersion \citep{snellen+2013,rodler+2014,benami+2018}.

Over the last decade, ground-based, high-dispersion (${\rp = \lambda/\Delta \lambda \sim 10^{5}}$) spectroscopy has provided robust detections of several molecular \citep[e.g.,][]{snellen+2010,birkby+2013,nugroho+2017} and atomic \citep[e.g.,][]{hoeijmakers+2018} species in giant exoplanet atmospheres. It uses cross-correlation of model templates with spectroscopic observations to probe the exoplanet's atmospheric composition. At these high spectral resolutions, molecular bands resolve into individual lines, whose signal contributions are co-added during cross-correlation. This technique is expected to be important in the ground-based detection of exoplanet biosignatures like oxygen, since the resolved telluric and exoplanet lines can be separated using the radial velocity of the system, the barycentric motion of the Earth, and the radial velocity component of the orbital motion of the exoplanet. Particularly amenable to this technique is the A-band transition of \oxy{} at 7600 \AA{} due to its spectral isolation and resolvability at high \rp{}.

\citet{snellen+2013} and \citet{rodler+2014} previously studied the feasibility of using ELT transmission spectroscopy to detect the \oxy{} A-band in terrestrial exoplanet atmospheres. \citet{snellen+2013} compare the detection requirements for an Earth-twin transiting in the habitable zones of G-dwarfs (G0--G5), early-type M-dwarfs (M0--M2), and late-type M-dwarfs (M4--M6), and find that the relatively short, 12-day periods make temperate exoplanets transiting late-type M-dwarfs preferential targets. Simulating European ELT (E-ELT) observations of an ${I=11.1\,\Mag}$\footnote{$I$-band magnitude expected for the brightest late-type M-dwarfs with a transiting, habitable-zone Earth-twin, assuming all late-type M-dwarfs host such a planet \citep{snellen+2013}.} Earth--M5V system assuming only uncorrelated noise, they determine a 3.8$\sigma$ detection of the \oxy{} A-band requires 30 transits. \citet{rodler+2014} perform similar simulations for various spectrograph setups, with and without uncorrelated noise. Adopting similar system parameters as \citet{snellen+2013}, for late-type M-dwarfs they determine 42--60 transits are needed for a 3$\sigma$ detection of \oxy{} when only uncorrelated noise is considered, and 60--84 transits when both uncorrelated and correlated noise are simulated.

The studies described above are based on idealized simulations with synthetic data. In particular, real data from real telescopes of real stars, which may show significant astrophysical noise, could potentially degrade the expected performances and affect the feasibility of such observations. In this paper, we test these feasibility predictions using real data. While high-dispersion E-ELT spectra of an ${I=11\,\Mag}$ M5V star will of course not be available for several years, they should be very similar to data of an ${I=7.5\,\Mag}$ dwarf from an 8m-class telescope. The fact that the light-gathering power of such a telescope is a factor $\sim$25 smaller is compensated by the star being $\sim$25 times brighter. We use three nights of high-dispersion, archival spectra from ESO's Very Large Telescope (VLT) of Proxima Centauri \citep[M5.0V, ${I=7.41\,\Mag}$;][]{jao+2014} and inject synthetic oxygen transmission spectra to assess the detectability of this biosignature gas in temperate terrestrial exoplanets. This multi-night spectral time series allows us to include the influence of the rich stellar spectra, the barycentric motion of the observatory, and all components of the Earth's atmosphere in our assessment of the recoverability of the exoplanet signal. 

The archival data and initial reduction are presented in Section \ref{sec:data}, and the simulation methodology in Section \ref{sec:methods}. The results are presented in Section \ref{sec:results} and discussed in Section \ref{sec:discussion}.

\section{Archival data of Proxima Centauri} \label{sec:data}

The dataset consists of three nights\footnote{Based on observations collected at the European Southern Observatory under ESO program 082.D-0953(A). PI: Liefke.} of archival observations of Proxima by the Ultraviolet and Visual Echelle Spectrograph (UVES, \citealp{dekker+2000}) mounted on UT2 of the VLT, taken 2009 March 10, 12, and 14. Each night of observation spans nearly eight hours, and consists of 215, 168, and 178 spectra respectively, with typical exposure times of 100 seconds for March 10 and 14, and between 90 and 200 seconds for March 12. The UVES slit width was $1''$ with a reported resolving power $\rp{} = 42,310$, corresponding to a spectral resolution of $\dl=0.18\,\mathrm{\AA}$ ($\dv = 7.1\,\kms$) at 7600 \AA{}, and a pixel sampling of 0.039 \AA{} ($\dv = 1.5\,\kms$). We inspected and removed spectral observations with poor \snr{}, leaving 214, 162, and 175 observations for each night.

\begin{figure}[t!]
    \includegraphics[height=0.35\textwidth]{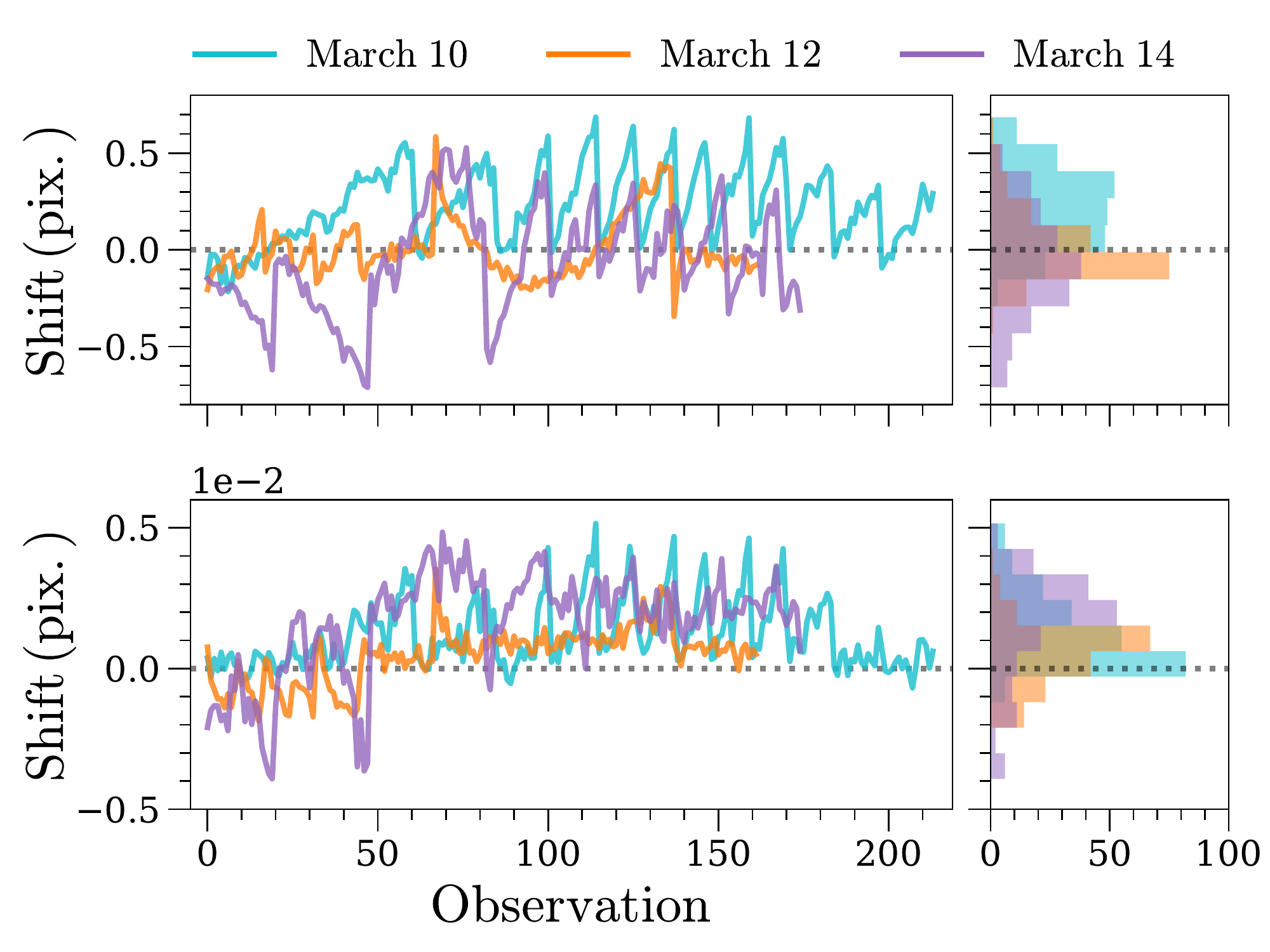}
	\caption{Left panels: Shifts in the wavelength solution in pixels (0.039 \AA{}, ${\dv = 1.5\,\kms}$) for the March 10 (blue), 12 (orange), and 14 (purple) spectral time series relative to the lowest-resolution reference spectrum for each night. The top and bottom panels correspond to the measured shifts before and after the SVD correction. Note the different vertical scales, showing that the shifts have decreased by two orders of magnitude.  Right panels: Corresponding histograms of wavelength shifts for each night. \label{fig:pixelShift}}
\end{figure}

\vspace{2cm}

\subsection{Initial Data Reduction} \label{sec:dataCorrection}

Initial reduction of the UVES data was performed by the UVES-Echelle pipeline\footnote{\path{https://www.eso.org/observing/dfo/quality/UVES/pipeline/pipe_reduc.html} ; \path{https://www.eso.org/observing/dfo/quality/UVES/pipeline/recipe_science.html}}, which includes de-biasing, background subtraction, order extraction, flat-fielding, wavelength calibration, and order merging. Upon inspection, variations in the wavelength solution over the course of each night are clearly visible in the pipeline-reduced spectra. These wavelength shifts have magnitudes of up to several tenths of a pixel (see Figure \ref{fig:pixelShift}), determined by cross-correlating the telluric-free region at 7500--7570 \AA{} of each spectrum with that of the lowest-resolution observation of each night. Additionally, we found a temporal variation in spectral resolution.

We used the singular value decomposition (SVD) method presented in \citet{rucinski1999} to simultaneously correct the wavelength shifts and variations in spectral resolution. In brief, SVD inverts the expression $F(\lambda) = B(\lambda) \ast f(\lambda)$ to calculate the kernel $B$ that broadens a narrow spectrum $f$ to $F$. For each night of UVES observations, we selected the spectrum with the lowest resolution to serve as reference and used the SVD method to derive 11-pixel-wide kernels for the remaining spectra based on the telluric-free 7500--7570 \AA{} wavelength range.
Convolving the spectra with their corresponding kernels corrects the variation in resolution. As Figure \ref{fig:pixelShift} demonstrates, the non-symmetric nature of the kernels also corrects the wavelength shifts, which have magnitudes on the order of thousandths of a pixel after convolution---a two order-of-magnitude improvement.

\subsection{Assessment of Data Quality} \label{sec:dataQuality}

\begin{figure}[t!]
	\includegraphics[height=0.35\textwidth]{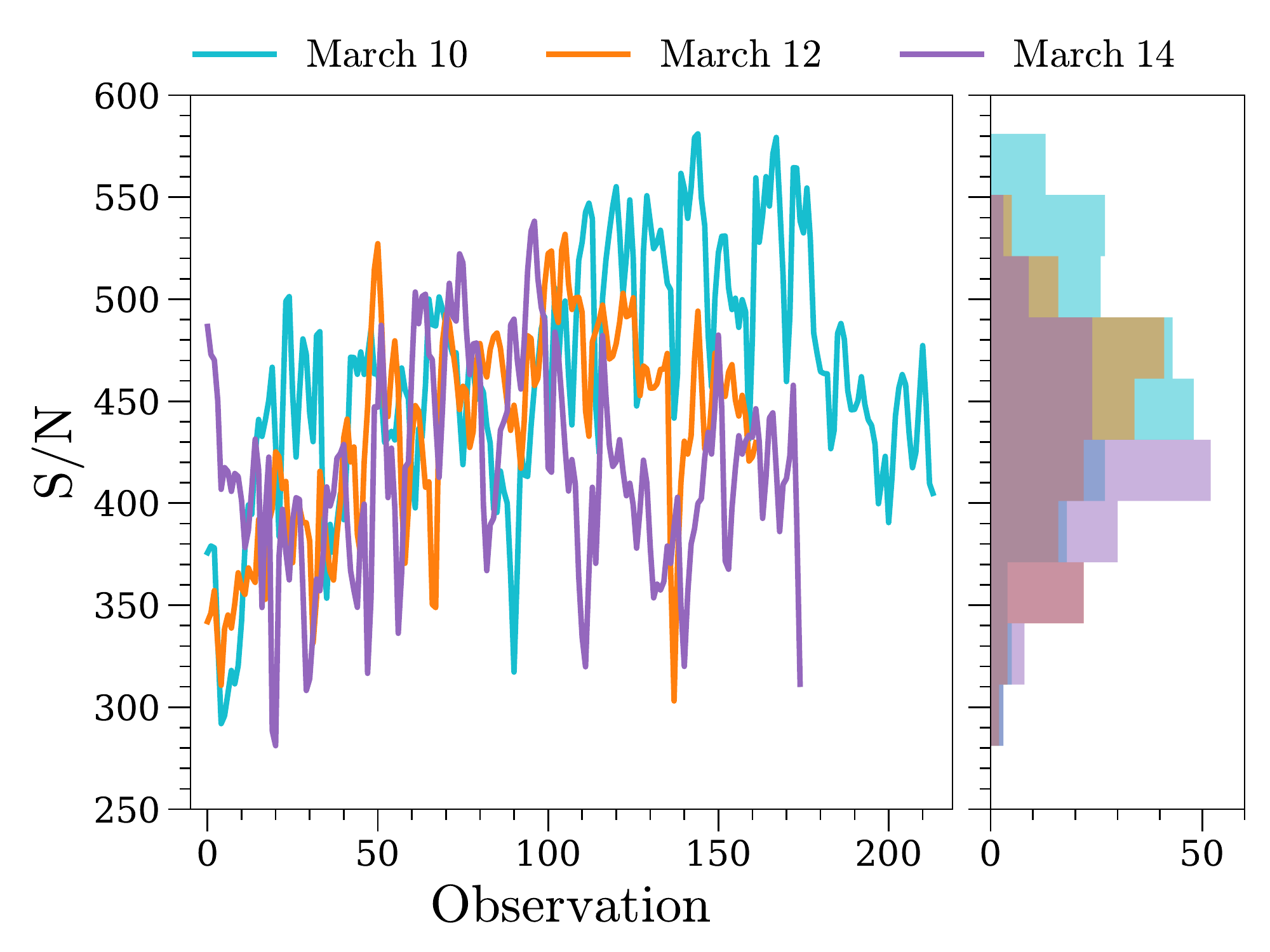}
	\caption{Left panel: \snr{} of each UVES spectrum. The observations are temporally sequenced and color-coded by date as in Figure \ref{fig:pixelShift}. Right panel: Histogram of \snr{} values for each observation date. The joint \snr{} distribution including all dates has a median value $443\pm57$.}
	\label{fig:snrVsObs}
\end{figure}

\begin{figure}[t!]
	\includegraphics[height=0.35\textwidth]{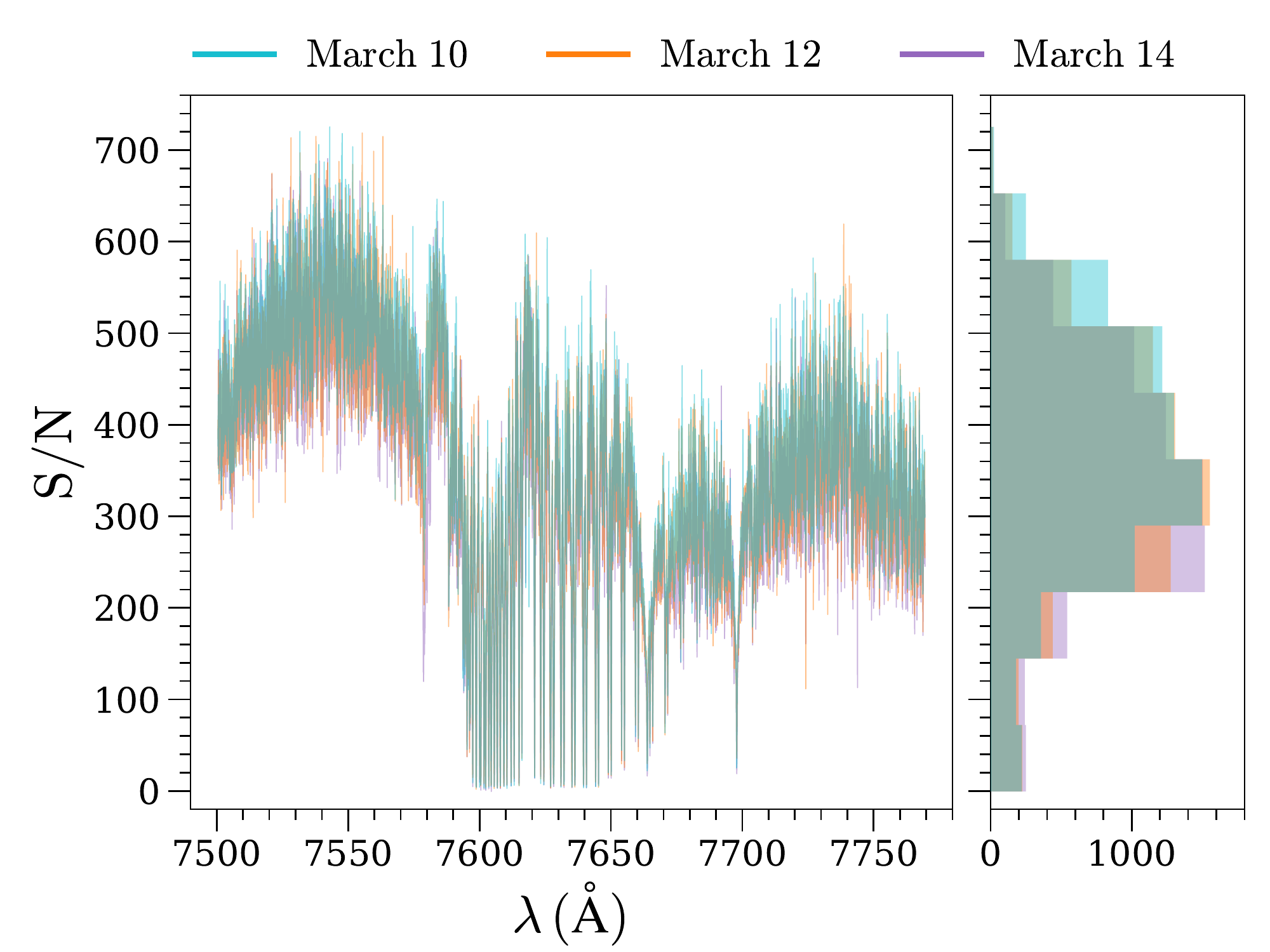}
	\caption{Left panel: \snr{} of each UVES wavelength bin, color-coded by observation date as in Figure \ref{fig:pixelShift}. Right panel: Histogram of \snr{} values for each observation date.}
	\label{fig:snrVsWave}
\end{figure}

Following the SVD correction, we calculated the \snr{} of the individual UVES spectra in the same telluric-free range by dividing $\sqrt{2}$ by the measured standard deviation of the ratio of two successive spectra. The average is taken of the \snr{} derived from both the preceding and subsequent spectra, resulting in
\begin{equation}
    (\snr{})_i = \frac{1}{\sqrt{2}} \left( \frac{1}{\mathrm{std}(f_i/f_{i-1})} + \frac{1}{\mathrm{std}(f_i/f_{i+1})} \right).
\end{equation}
Figure \ref{fig:snrVsObs} shows the temporal evolution of the \snr{} over the course of each night, and the associated distributions. Overall, the spectra have a median \snr{} of $443\pm57$ in the telluric-free region. We also calculated the \snr{} for each UVES wavelength bin by dividing the bin's median flux value by its corresponding standard deviation. These \snr{} profiles are shown in Figure \ref{fig:snrVsWave} for each observation night.

An important property of the archival observations is the total throughput of the telescope and instrument. Based on the $I$-band magnitude of Proxima, the collecting area of the VLT, and the integration times, we determined the incident photon count in the telluric-free range and compared this to the observed UVES photon count to find a throughput of 3.8\%, a factor $\sim$2.5 lower than the new ESPRESSO spectrograph (C. Lovis, private comm.), and $\sim$5 times lower than the throughput assumed by \citet{snellen+2013}.

\vspace{0.5cm}

\section{Synthetic oxygen transmission spectra} \label{sec:methods}

\subsection{Transmission Model Calculation and Injection} \label{sec:model}

\begin{figure}[t!]
	\includegraphics[height=0.35\textwidth]{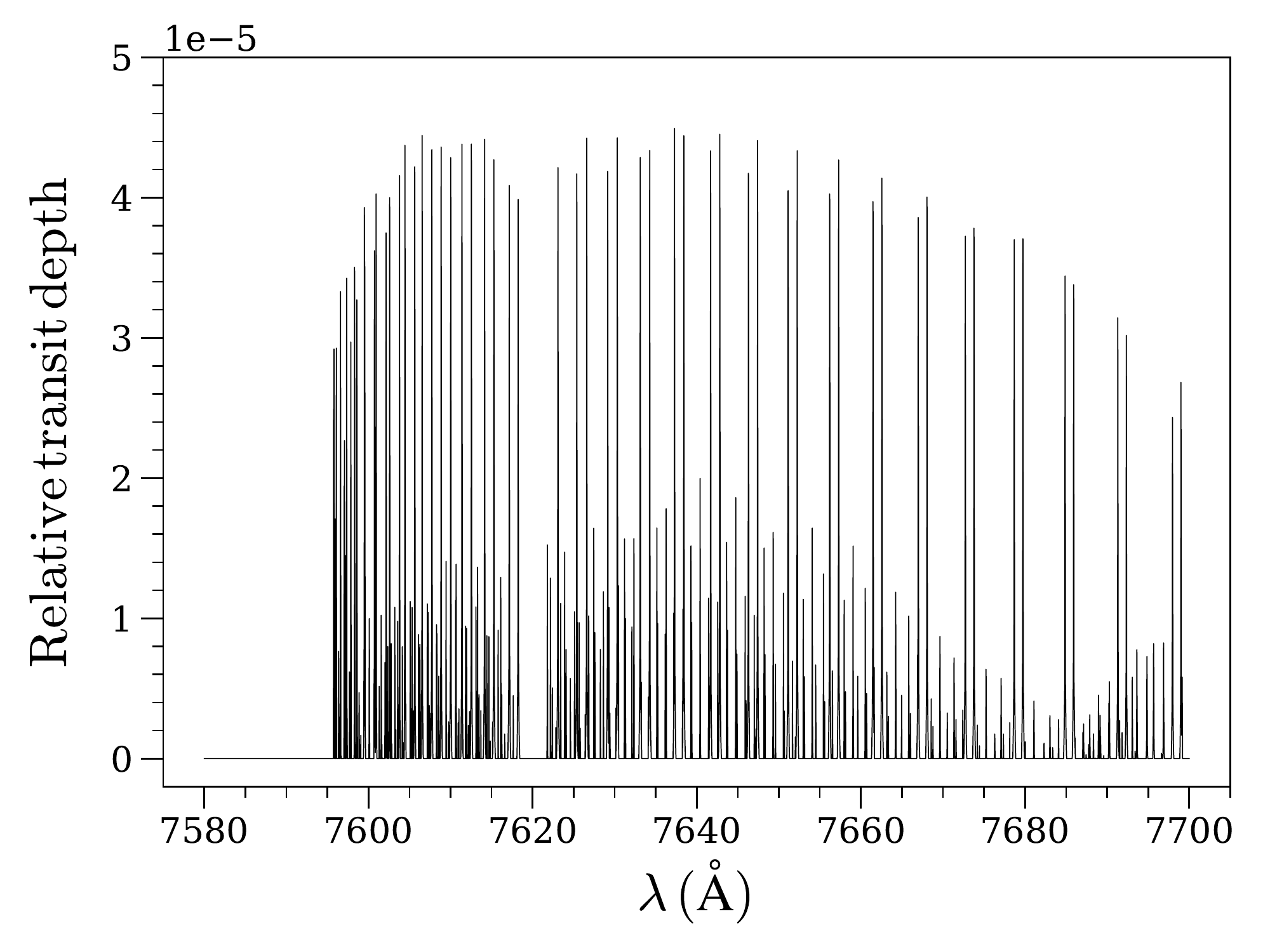}
	\caption{High-resolution (\rp{} = 100,000) model transmission spectrum of the \oxy{} A-band for an Earth-twin transiting an M5V star. A high-pass filter was used to remove the broadband component in the transmission spectrum for which the high-dispersion transit technique is insensitive. 
	\label{fig:o2Model}}
\end{figure}

We determined the atmospheric transmission in the oxygen A-band at a spectral resolving power of ${\rp{} =100,000}$, and the resulting change in effective radius as a function of wavelength, for an Earth twin---a planet with Earth's surface gravity, and atmospheric composition and temperature structure (including a constant oxygen volume-mixing ratio of 20.95\%). This was converted to a transmission spectrum of an Earth-twin transiting a late-type M-dwarf assuming an M5V star radius of 0.19 \rsun. Since the high-dispersion transit technique is only sensitive to the high-frequency part of the transmission spectrum, the broadband transmission signal is removed using a high-pass filter. Figure \ref{fig:o2Model} shows the resulting, relative transit depth due to \oxy{}, with individual line strengths of up to $4.5 \times 10^{-5}$. Refraction effects were not modeled as they negligibly impact the transmission spectrum of an Earth--M5V system \citep{betremieux+2014}.

We simulate the observations of a full transit by injecting the model template (Figure \ref{fig:o2Model}) into a series of UVES observations, Doppler-shifted to mimic the orbital motion of the exoplanet about its system's barycenter. Assuming a circular orbit and an edge-on inclination, the transit duration for an Earth-twin orbiting an M5V star in the habitable zone is 1.4 hours for an orbital period $P=11.8\,\mathrm{days}$, stellar mass $\mstar = 0.19 \msun$, and stellar radius $\rstar=0.19\rsun$ \citep{snellen+2013}. Adopting a uniform cadence of 130 seconds (median observation cadence of the UVES data), each simulated transit consists of 39 successive UVES observations injected with an Earth-twin \oxy{} signal. For a given radial velocity offset (see Section \ref{sec:singleV}), we are able to simulate 13 transits using the UVES time-series in its original order, without the need to re-use or re-order spectra. 

\subsection{Signal Recovery} \label{sec:UVESsim}

Subsequently, to retrieve the \oxy{} transmission signal, we normalize each spectrum to its median flux value on the range 7500--7570 \AA{}, and flag regions where overlapping spectral orders are poorly stitched. We also flag the stellar potassium doublet at 7665/7699 \AA{}, as well as the most-saturated telluric \oxy{} lines.  Several steps are required to remove the telluric \oxy{} and (weak) water transmission spectrum. The measured fluxes at each wavelength step are first divided by their median value, essentially normalizing each wavelength column to its typical depth for that night. We subsequently fit each column with a quadratic function in airmass, and divide out the main temporal variations. Finally, we use singular value decomposition to identify and remove the strongest residual noise components present in the dataset, as used by \cite{kok+2013}. The number of SVD components is chosen such that the \snr{} from combining the 13 unique transits at a given offset velocity is maximized (see Section \ref{sec:singleV}). A high-pass filter is subsequently applied to remove low-frequency trends in the spectra.

The exoplanet \oxy{} signal is extracted by cross-correlating each filtered transit spectrum with the \oxy{} template spectrum over velocities ranging from $-100$ to $+100$ \kms{} in steps of 1 \kms{} in the exoplanet rest-frame. This results in 39 cross-correlation functions (\ccf{}s) per transit, of which the sum corresponds to its overall transmission signal. We perform the same analysis on the spectra without injected transmission signals to assess the retrieved \snr{} of \oxy{}, by comparing the cross-correlation function of the spectra with and without injected signals, using
\begin{equation}
\snr{} = \frac{ \mathrm{max}( \ccf{inj}-\ccf{0} ) }{ \mathrm{std}( \ccf{0} ) },
\end{equation}
where $\ccf{inj}$ and \ccf{0} are the summed cross-correlation functions for the injected and non-injected spectra, and $\mathrm{std}( \ccf{0} )$ is the standard deviation of the summed CCF of the non-injected spectra.

\subsection{Perfect Gaussian-Noise Comparisons} \label{sec:GaussSim}

We also repeat each transit simulation using synthetic data consisting of pure Gaussian noise to quantify the influence of systematic effects on the signal retrieval. For each UVES observation we create a corresponding synthetic ``spectrum" by sampling a Gaussian normal distribution on a pixel-by-pixel basis, with a standard deviation such that it matches the \snr{} of the spectrum in the telluric-free region scaled by the square-root of the relative flux.

A nearly similar procedure is followed for the synthetic dataset as for the real spectra to retrieve the oxygen transmission signals. Continuum normalization, wavelength-column normalization, and airmass and SVD corrections are not necessary since there are, by definition, no systematic effects in the synthetic data. We do perform the same flagging procedures and filtering to preserve their effect on the \ccf{}.

\section{Results} \label{sec:results}

\subsection{Transits at Constant \voff{}} \label{sec:singleV}

\begin{figure}[t!]
	\includegraphics[height=0.35\textwidth]{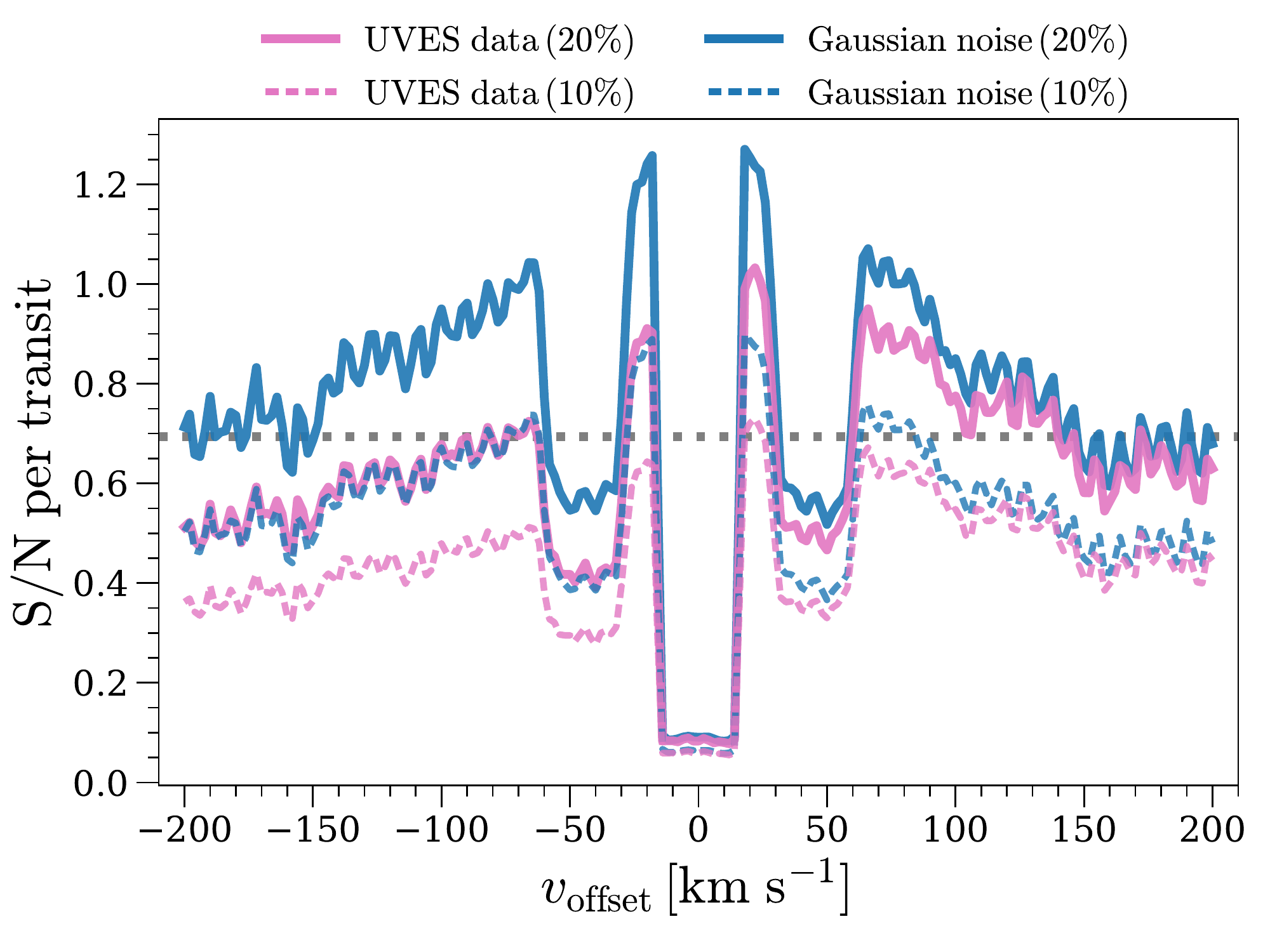}
	\caption{Average signal-to-noise per transit for molecular oxygen as a function of velocity offset, which incorporates both the systematic velocity of the target and the radial component of the Earth's barycentric velocity. The purple lines denote the UVES data results while the blue lines indicate the median Gaussian-noise results for instrument+telescope throughputs of 20\% (solid) and 10\% (dashed). The dotted, grey line shows the prediction of \citet{snellen+2013}, who assumed a 20\% throughput.}
	 \label{fig:singleV}
\end{figure}

We inject 13 transits for velocity offsets \voff{} ranging from -200 to +200 \kms{} in steps of 2 \kms{} in the UVES data of Proxima, using the methods outlined in Section \ref{sec:methods}. The \voff{} factor combines both the systematic velocity of the target and the radial component of the Earth's barycentric orbital velocity, which we take as constant since the induced shift has a magnitude $\sim$0.01 pixels across each 8-hour set of observations. We also neglect the Earth's rotation, since this effect only induces a $\sim$0.1 pixel shift over 8 hours. Since we use real data, the stellar lines do shift by these amounts during the observations, however our analysis removes their effect. By simulating transits at different \voff{} we can mimic observations at different times of year.

The average, per-transit \snr{} for each offset velocity at an instrument+telescope throughput of 20\%, calculated from the 13 unique transits at that \voff{}, is shown in solid purple in Figure \ref{fig:singleV}. The \snr{} depends on the overlap with telluric lines, which depends on \voff{}, and ranges from \snr{} = 0.4 to 1.0 for $|\voff{}|>16\,\kms{}$. At $|\voff{}|<16\,\kms{}$, the injected exoplanet \oxy{} lines fall within the flagging bounds for the heavily-saturated telluric \oxy{} lines, leading to essentially no retrieved exoplanet signal. Since the strongest \oxy{} A-band lines lie in pairs, maximum \snr{} occurs around $\voff{} = \pm 22\,\kms{}$ when both exoplanet lines are free of tellurics. As discussed in Section \ref{sec:GaussSim}, we also perform transit simulations using synthetic data consisting of purely Gaussian noise. For each \voff{} we repeat the 13 transit simulations for 100 initializations of Gaussian-noise data. The median per-transit \snr{} for each \voff{} at a throughput of 20\% is shown in Figure \ref{fig:singleV} in solid blue, and mimics the trend seen for the UVES simulations, albeit up to 45\% higher. The per-transit \snr{} values for both the UVES and Gaussian simulations are generally consistent with the per-transit \snr{} predicted by \cite{snellen+2013}.

\vspace{0.5cm}

\subsection{Combining Transits at Multiple \voff{}} \label{sec:multiV}

\begin{figure}[t!]
	\includegraphics[height=0.35\textwidth]{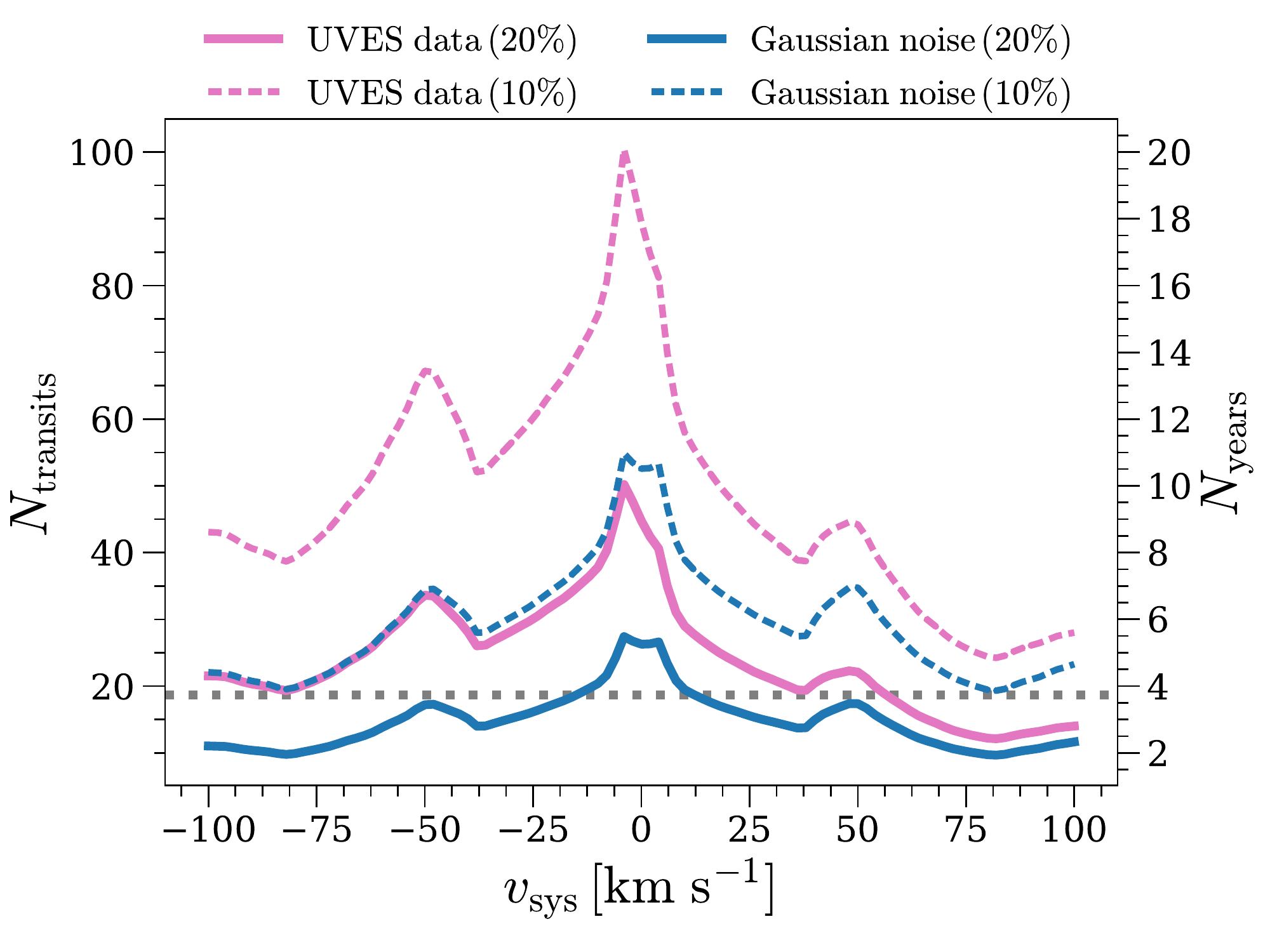}
	\caption{Left axis: Estimate of the number of transits required to achieve a 3$\sigma$ molecular oxygen detection as a function of the systematic velocity of the target, accounting for the change in barycentric velocity due to Earth's motion around the Sun. Right axis: Corresponding number of years needed to collect the transit observations, assuming a typical observability. The estimates based on UVES data and Gaussian noise are shown in purple and blue, respectively, for 20\% (solid) and 10\% (dashed) instrument+telescope throughputs. The \cite{snellen+2013} prediction assumed a 20\% throughput, and is denoted in dotted grey.}
	\label{fig:multiV}
\end{figure}

To estimate the number of transits required to detect \oxy{}, we vary \voff{} to mimic the change in the radial component of Earth's orbital motion towards the star during an observing season. For each systematic velocity \vsys{}, the $N$ per-transit \snr{} values from Figure \ref{fig:singleV} that fall within \vsys{} $\pm$ 20 \kms{} are added in quadrature, and then normalized by $\sqrt{N}$. This results in the nominal \snr{} per transit expected during an observing season centered at \vsys{}. This is subsequently scaled to estimate how many transits observations are required to reach a 3$\sigma$ \oxy{} signal.

For \vsys{} ranging from -100 to +100 \kms{} in steps of 2 \kms{}, Figure \ref{fig:multiV} plots the number of transits required to reach this level, as well as the corresponding number of years to collect such observations, assuming five\footnote{Targets are typically above the horizon 50\% of the time, and only 1/3 of the time will it be sufficiently dark at the observatory. Assuming an orbital period of 11.8 days, this results in five transits per year.} transits are observable each year. The UVES data indicate that the required number of transits to achieve an \oxy{} detection of $\snr{} = 3$ for an instrument+telescope throughput of 20\% ranges from 20 to 50 for \vsys{} of 0 to $\pm$50 \kms{}, corresponding to 4 to 10 years of observations. Note that the total amount of observing time needed will range between 70 and 175 hours over this period of time.

\section{Discussion and Conclusions} \label{sec:discussion}

The number of transits required for an \oxy{} detection, as predicted by using real spectroscopic data of Proxima, is very similar to that predicted by \cite{snellen+2013} and \cite{rodler+2014}. This implies that real astronomical and instrumental effects that were not considered in the previous simulations do not strongly affect the power of the cross-correlation technique in retrieving molecular oxygen. In particular, the archival UVES spectra show considerable wavelength instabilities which have been effectively removed by the methods presented in Section \ref{sec:dataCorrection}. In addition, the strong \snr{} variations as a function of wavelength due to the dense forest of stellar molecular lines and saturated telluric oxygen lines were not prohibitive in recovering the exoplanet signal over a wide range of radial velocities. As expected, any micro-tellurics are effectively taken out by the singular value decomposition (Section \ref{sec:UVESsim}). During the three nights of observation, the occurrence of variable line emission in the cores of the 7665/7699 \AA{} stellar potassium doublet shows that Proxima experienced several flares. Our masking of the potassium doublet line cores (Section \ref{sec:UVESsim}) was sufficient to mitigate their potential effect on the oxygen retrieval.

Although we injected the exoplanet \oxy{} signal at ${\rp=10^{5}}$, the UVES data have $\rp \sim 40,000$. We do not expect performing this study using $\rp \sim 10^5$ tellurics will substantially improve our results in terms of the range of \voff{} that can be probed. The strongest telluric \oxy{} lines we flag are saturated, so increasing the resolving power will not dramatically narrow their line widths.

We conclude, using archival UVES data of Proxima, that a few dozen transits observed with the future ELTs are required to detect molecular oxygen from an Earth twin transiting an ${I=11\,\Mag{}}$ M5V star, assuming an instrument+telescope throughput of 20\% and a resolving power of $\rp{}=100,000$.  For a single ELT, the required number of transits can be collected on a time scale of 4 to 10 years, very similar to that predicted by \cite{snellen+2013} and \cite{rodler+2014}.

For this method to live up to its potential, a high instrumental throughput is key, since the required number of transits is linearly dependent on it. While the newest generation of high-dispersion spectrographs can achieve throughputs in the range 10--20\%, we emphasize the importance of further development of instrument design to increase the throughput, e.g., through novel designs of high-dispersion spectrographs that specifically target the molecular oxygen band(s) \citep{benami+2018}.

Additionally, we note strong instrumental effects present in the UVES data, particularly those affecting the wavelength solution, require us to implement SVD techniques to mitigate their influence on the transmission signal. These techniques are known to remove part of the exoplanet signal. Indeed, we determine signal losses $\sim$10--20\% for $|\voff{}| > 16 \, \kms{}$. A stabilized spectrograph would not suffer from such instrumental effects, which may improve the \snr{} per transit.

\acknowledgments
The authors acknowledge support from the European Research Council under the European Union's Horizon 2020 research and innovation programme under grant agreement No. 694513. The authors thank the anonymous referee for their insightful comments. The authors also thank R. J. de Kok for use of his \oxy{} transmission models, and F. J. Alonso-Floriano, O. Contigiani, H. J. Hoeijmakers, D. J. M. Petit dit de la Roche, and A. R. Ridden-Harper for helpful discussions.

\vspace{5mm}
\facilities{VLT(UVES)}


\begin{thebibliography}{}

\bibitem[Ben-Ami et al.(2018)]{benami+2018} Ben-Ami, S., L{\'o}pez-Morales, M., Garcia-Mejia, J., Gonzalez Abad, G., \& Szentgyorgyi, A.\ 2018, \apj, 861, 79

\bibitem[B{\'e}tr{\'e}mieux \& Kaltenegger(2014)]{betremieux+2014} B{\'e}tr{\'e}mieux, Y., \& Kaltenegger, L.\ 2014, \apj, 791, 7

\bibitem[Birkby et al.(2013)]{birkby+2013} Birkby, J.~L., de Kok, R.~J., Brogi, M., et al.\ 2013, \mnras, 436, L35

\bibitem[de Kok et al.(2013)]{kok+2013} de Kok, R.~J., Brogi, M., Snellen, I.~A.~G., et al.\ 2013, \aap, 554, A82

\bibitem[Dekker et al.(2000)]{dekker+2000} Dekker, H., D'Odorico, S., Kaufer, A., Delabre, B., \& Kotzlowski, H.\ 2000, \procspie, 4008, 534

\bibitem[Hoeijmakers et al.(2018)]{hoeijmakers+2018} Hoeijmakers, H.~J., Ehrenreich, D., Heng, K., et al.\ 2018, \nat, 560, 453

\bibitem[Jao et al.(2014)]{jao+2014} Jao, W.-C., Henry, T.~J., Subasavage, J.~P., et al.\ 2014, \aj, 147, 21

\bibitem[Lippincott et al.(1967)]{lippincott+1967} Lippincott, E.~R., Eck, R.~V., Dayhoff, M.~O., \& Sagan, C.\ 1967, \apj, 147, 753

\bibitem[Lovelock(1965)]{lovelock1965} Lovelock, J.~E.\ 1965, \nat, 207, 568

\bibitem[Lovelock(1975)]{lovelock1975} Lovelock, J.~E.\ 1975, Proceedings of the Royal Society of London Series B, 189, 167

\bibitem[Meadows(2017)]{meadows2017} Meadows, V.~S.\ 2017, Astrobiology, 17, 1022

\bibitem[Meadows et al.(2018)]{meadows+2018} Meadows, V.~S., Reinhard, C.~T., Arney, G.~N., et al.\ 2018, Astrobiology, 18, 630

\bibitem[Nugroho et al.(2017)]{nugroho+2017} Nugroho, S.~K., Kawahara, H., Masuda, K., et al.\ 2017, \aj, 154, 221

\bibitem[Rodler \& L{\'o}pez-Morales(2014)]{rodler+2014} Rodler, F., \& L{\'o}pez-Morales, M.\ 2014, \apj, 781, 54

\bibitem[Rucinski(1999)]{rucinski1999} Rucinski, S.\ 1999, IAU Colloq.~170: Precise Stellar Radial Velocities, 185, 82

\bibitem[Sagan et al.(1993)]{sagan+1993} Sagan, C., Thompson, W.~R., Carlson, R., Gurnett, D., \& Hord, C.\ 1993, \nat, 365, 715

\bibitem[Segura et al.(2005)]{segura+2005} Segura, A., Kasting, J.~F., Meadows, V., et al.\ 2005, Astrobiology, 5, 706

\bibitem[Snellen et al.(2010)]{snellen+2010} Snellen, I.~A.~G., de Kok, R.~J., de Mooij, E.~J.~W., \& Albrecht, S.\ 2010, \nat, 465, 1049

\bibitem[Snellen et al.(2013)]{snellen+2013} Snellen, I.~A.~G., de Kok, R.~J., le Poole, R., Brogi, M., \& Birkby, J.\ 2013, \apj, 764, 182

\end{thebibliography}
\end{document}